\documentclass[fleqn,usenatbib]{mnras}

\usepackage{newtxtext,newtxmath}

\usepackage[T1]{fontenc}
\usepackage{ae,aecompl}

\usepackage{graphicx}	
\usepackage{amsmath}	
\usepackage{amssymb}	
\usepackage{multicol}        
\usepackage{bm}		
\usepackage{pdflscape}	

\title[Molecular hydrogen formation on interstellar PAHs]{Molecular hydrogen formation on interstellar PAHs through Eley-Rideal abstraction reactions}

\author[N. Foley et al.]{Nolan Foley$^{1}$, S. Cazaux$^{2,3}$, D. Egorov$^{1}$, L.M.P.V. Boschman$^{1,4}$, R. Hoekstra$^{1}$ \
\newauthor and T. Schlath\"olter$^{1}$ \thanks{Contact e-mail: t.a.schlatholter@rug.nl}
\\
$^{1}$Zernike Institute for Advanced Materials, University of Groningen, Nijenborgh 4, 9747 AG Groningen, The Netherlands\\
$^{2}$Faculty of Aerospace Engineering, Delft University of Technology, Kluyverweg 1, 2629 HS, Delft, The Netherlands\\
$^{3}$University of Leiden, P.O. Box 9513, NL, 2300 RA, Leiden, The Netherlands.\\
$^{4}$Kapteyn Astronomical Institute, University of Groningen, Landleven 12, 9747 AD, Groningen, The Netherlands
}

\date{Accepted XXX. Received YYY; in original form ZZZ}

\pubyear{2018}

\begin{document}
\label{firstpage}
\pagerange{\pageref{firstpage}--\pageref{lastpage}}
\maketitle

\begin{abstract}
We present experimental data on H$_2$ formation processes on gas-phase polycyclic aromatic hydrocarbon (PAH) cations. This process was studied by exposing coronene radical cations, confined in a radio-frequency ion trap, to gas phase H atoms. Sequential attachment of up to 23 hydrogen atoms has been observed.  Exposure to atomic D instead of H allows one to distinguish attachment from competing  abstraction reactions, as the latter now leave a unique fingerprint in the measured mass spectra. Modeling of the experimental results using realistic cross sections and barriers for attachment and abstraction yield a 1:2 ratio of abstraction to attachment cross sections. The strong contribution of abstraction indicates that H$_2$ formation on interstellar PAH cations is an order of magnitude more relevant than previously thought.
\end{abstract}


\section{Introduction}
The formation of molecular hydrogen in the interstellar medium (ISM) is a topic of ongoing debate (see recent review \cite{wakelam2017}). As molecular hydrogen is the most abundant molecule in the universe and plays a key role in many astrophysical processes, a proper understanding of the processes that lead to its formation is of great interest. A number of potential formation processes have already been explored. While gas phase routes to H$_2$ formation have been found to be inefficient, formation on interstellar dust grains has been identified as one possible mechanism \citep{oort1946,gould1963}. Another possible route to molecular hydrogen formation is on interstellar polycyclic aromatic hydrocarbons (PAHs) \citep{bauschlicher1998,hirama2004,lepage2009,mennella2012,boschman2012,thrower2012}. A variety of objects inside and outside our galaxy exhibit spectra crowded with unidentified lines. These lines can be seen in absorption in the optical (diffuse interstellar bands: DIBS) and in emission in the infrared (aromatic infrared bands: AIB). Ubiquitous interstellar AIB are now commonly considered to be carried by PAHs. These PAHs are typically at least partially hydrogenated \citep{schutte1993,bernstein1996}. Assuming DIBS to be due to PAH-based species as well, it is most likely that part are present in protonated form or as syperhydrogenated cations, which feature the observed transitions in the visible \citep{lepage1997,snow1998,hammonds2009}. 

Atomic hydrogen impacting on PAH may undergo an attachment reaction and become attached to the molecule. Sequences of these addition reactions, with s additional H atoms, are of the type 
\begin{equation}
[\mbox{PAH}+s\mbox{H}]+\mbox{H}\rightarrow [\mbox{PAH}+(s+1)\mbox{H}], \;\; \Delta M=+1
\label{eq:H-attachment}
\end{equation}
\noindent
are referred to as hydrogenation sequences, and leave the PAH in a state of superhydrogenation and increase its mass $M$ by 1. Recent experimental and theoretical research on trapped coronene radical cations C$_{24}$H$_{12}^+$ revealed that in the gas-phase and at $T=300$ K, hydrogenation proceeds through a specific sequence of well-defined atomic sites. Reaction barriers and binding energies lead to odd - even oscillations in the observed superhydrogenation states, and magic numbers of particularly high intensity for the attachment of $n=$5, 11, and 17 extra H atoms \citep{boschman2012,cazaux2016}.

Superhydrogenation dramatically alters the response of neutral and ionic gas-phase PAHs in various astrochemically relevant interaction processes. 
Attachment of small numbers of H atoms to coronene cations can for instance quench photoionization-induced H loss from a C$_{24}$H$_{12}^+$ precursor cation \citep{reitsma2014,reitsma2015}. For superhydrogenation of neutral pyrene molecules,
an opposite effect was observed.The C-backbone is weakened and fragmentation upon ion collisions or photoionization is increased \citep{wolf2016,gatchell2015}. Attachment of a single H atom to a PAH radical cations has a dramatic influence on its IR spectrum \citep{knorke2009} and substantially decreases the HOMO-LUMO gap \citep{pathak2008}.

\begin{figure*}
\includegraphics[width=170mm]{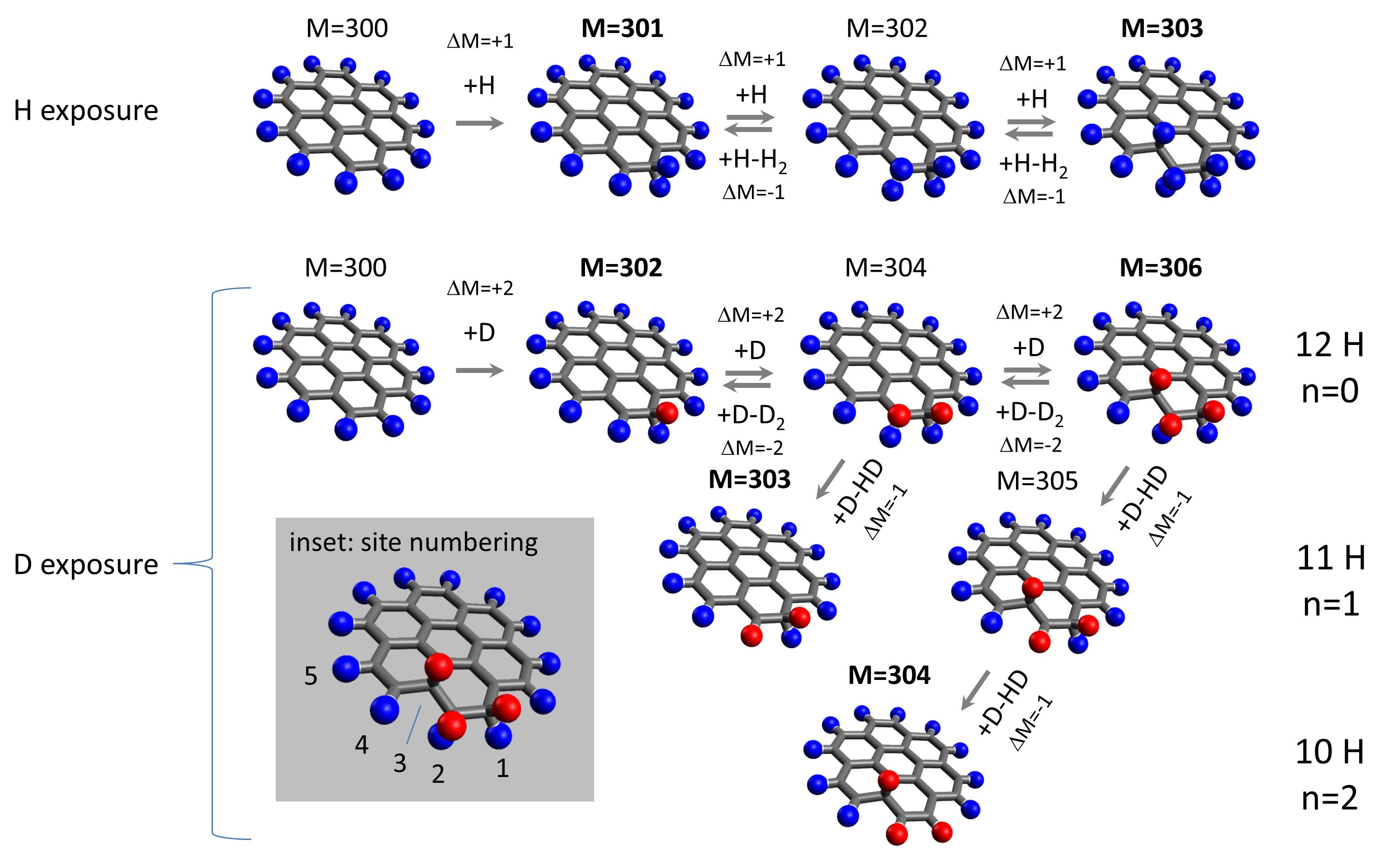}
\caption{Schematic representation of the reaction sequences for subsequent H (top row) or D (bottom row network) interactions with a coronene cation. H atoms are marked in blue and D atoms in red. Only a selection of possible attachment (+H, +D) and abstraction (+H-H$_2$, +D-D$_2$, +D-HD) processes are indicated. The third atom is attached to an inner edge site, as indicated by recent IR spectroscopy data \citep{schlatholter2018} and not an outer edge site as previously predicted \citep{cazaux2016}. The masses of systems that possess an odd total number of H+D atoms are given in bold letters.}
\label{fig:sketch}
\end{figure*}

In 1998 a second type of H reaction with PAH molecules was proposed \citep{bauschlicher1998}, which was experimentally confirmed in 2012 through hydrogenation experiments on supported PAH thin films by Mennella et al. \citep{mennella2012}. In these so-called direct abstraction reactions of the Eley-Rideal type   
\begin{equation}
[\mbox{PAH}+s\mbox{H}]+\mbox{H}\rightarrow [\mbox{PAH}+(s-1)\mbox{H}]+\mbox{H}_2, \;\; \Delta M=-1   
\label{eq:H2-abstraction}
\end{equation}

\noindent
the incoming H atom does not get bound to the PAH molecule, but rather reacts with an H atom already present in the hydrogenated PAH, to directly desorb as an H$_2$ molecule. In these experiments on solid PAH films the abstraction channel is determined to be more than an order of magnitude weaker than H attachment, with the reaction cross sections for abstraction and attachment being respectively 0.06 \AA$^2$ and 1.1 \AA$^2$ corresponding to a ratio between abstraction and attachment of $\approx 1:20$  \citep{mennella2012}.

In the following, we study H abstraction reactions on gas phase coronene cations, C$_{24}$H$_{12}^+$. It should be noted that coronene is not a major species in the interstellar medium \citep{hirama2004}, but it is used as one of the prototypical PAHs in related astrolaboratory research \citep{boschman2012,cazaux2016,rauls2008,boschman2015,jochims1994,ling1998} because it is fairly large and has a compact shape, making it relatively easy to work with, and it is commercially available in large quantities. By comparing the mass spectra obtained from C$_{24}$H$_{12}^+$ exposure to $T=300$\;K H and D beams, respectively, direct evidence for the occurrence of abstraction reactions is observed. The modeling of the measured mass spectra with a time-dependent rate equation model indicates a relative cross section for abstraction that is about an order of magnitude larger than previously thought.

\section{Experiment}
\subsection{Concept of the experiment}
The hydrogenation of coronene cations is predominantly determined by the alternating heights of the energy barriers for hydrogen addition. Even numbered superhydrogenation states can be subject to barrierless hydrogenation, whereas hydrogen attachment to odd-numbered superhydrogenation states involves a reaction barrier. As a result, all (closed shell) odd-numbered superhydrogenation states are more stable than the (radical) even-numbered states and occur significantly more often. A typical mass spectrum consists of dominating peaks at $M=301$, 303, 305 etc.  On top of that, ''magic'' stages of superhydrogenation are observed for the attachment of 5, 11, and 17 H atoms, corresponding to hydrogenation stages which have particularly high binding energies \citep{cazaux2016}. 

In principle, an incoming H atom can interact with every C site in a coronene cation. As a first step, the  attachment on one of the outer edge positions is energetically most favorable and thus most likely \citep{mennella2012,cazaux2016}. As a result, two H atoms are attached to a single carbon atom, which we shall refer to as the position being doubly occupied. Figure 1 illustrates the initial hydrogenation and abstraction steps for a coronene cation. In the top row, the case of H exposure is sketched. The first H attachment leads to a doubly occupied outer edge site and results in a mass increase by one unit into $M=301$. Previous experimental studies have shown that this process quickly transfers the entire C$_{24}$H$_{12}^+$ population into C$_{24}$H$_{13}^+$ \citep{boschman2012}. This implies that the probability of abstraction from C$_{24}$H$_{13}^+$ is negligibly small. Attachment of a second H atom leads to double occupation of the adjacent outer edge site and a molecular mass of $M=302$. From here on further hydrogenation competes with abstraction, if an impinging H atom impacts on a previously created doubly occupied site and undergoes an Eley-Rideal reaction \citep{mennella2012} with one of the H atoms attached to the site. This leads to the release of a neutral H$_2$ molecule which corresponds to the net loss of an H atom by the molecule (see eq. 2). 

From a mass spectrometric perspective, it is important to realize that since hydrogenation and abstraction shift the mass of a superhydrogenated coronene cation by +1 and -1, respectively, it is not possible to establish whether a C$_{24}$H$_{12+n}^+$ originates from C$_{24}$H$_{12+(n-1)}^+$ via H attachment or from C$_{24}$H$_{12+(n+1)}^+$ via H abstraction. This explains why abstraction reactions remained obscured thus far in gas phase hydrogenation experiments. Furthermore, as abstraction counteracts the mass shift towards higher masses driven by hydrogenation, the rates for H addition might well be underestimated.

In order to overcome this problem and quantify the relative contribution of abstraction reactions on gas-phase coronene cations, the cations can be exposed to atomic D ($^2$H) rather than H ($^1$H). Addition and abstraction of atomic D then change the mass of the molecular precursor by +2 and -2, respectively. Except for the twice as large step size in mass, this yields a similar spectrum as for hydrogen. The doubly occupied sites, however, are initially occupied by an H and a D atom, i.e. the abstraction reaction can also involve an H atom, leading to a mass change of only -1. For our prototypical system, coronene with an initial mass of 300, the appearance of molecular ions with odd mass numbers is a direct signature of such an abstraction. More specific for a coronene cation C$_{24}$H$_{(12-n)}$\,D$_m^+$ that has lost $n$ of its initial H atoms by HD abstraction and contains $m$ additional D atoms the following generic reactions need to be considered: 

\begin{equation}
\mbox{C}_{24}\,\mbox{H}_{(12-n)}\,\mbox{D}_m^+ \; +\; \mbox{D} \; \longrightarrow \left \lbrace
\begin{array}{lr}
\mbox{C}_{24}\,\mbox{H}_{(12-n)}\,\mbox{D}_{(m+1)}^+ & \Delta M=+2 \\ \\
\mbox{C}_{24}\,\mbox{H}_{(12-n)}\,\mbox{D}_{(m-1)}^+ \;+\; \mbox{D}_2 & \Delta M=-2 \\ \\
\mbox{C}_{24}\,\mbox{H}_{(12-n-1)}\,\mbox{D}_m^+ \;+\; \mbox{HD} & \Delta M=-1
\end{array}
\right.
\label{eq:HD-abstraction}
\end{equation}

\noindent
The reaction network driven by thermal D exposure described by eq.\ref{eq:HD-abstraction} is schematically illustrated in the bottom part of Figure 1. As for H, it is assumed that the singly hydrogenated cation is not undergoing significant abstraction reactions. In the figure H atoms are marked in blue and D atoms in red. From the figure and eq. \ref{eq:HD-abstraction} it is clear that for attachment and abstraction of a D atom the systems changes mass in steps of $\pm 2$ and stays in the same row. Loss of an H atom (HD abstraction)  moves the molecular system one row down. The rows are characterized by the number of H atoms removed from the precursor coronene cation. The rows with odd numbers of H atoms abstracted correspond to odd-mass cations and are a direct signature of abstraction which allow us to determine both attachment and abstraction cross sections and barriers.

\subsection{Experimental implementation}
The coronene radical cations were produced by means of electrospray ionization (ESI) from a coronene solution in methanol. Admixture of AgNO$_3$ to the solution facilitates charge transfer from C$_{24}$H$_{12}$ to Ag$^+$. The C$_{24}$H$_{12}^+$ beam generated by the ESI source was then phase space compressed in a RF ion funnel and mass selected in a RF quadrupole mass filter. The ions were then trapped in a 3D RF ion trap at ambient temperature \citep{bari2011,egorov2016}. Note, that the substantial binding energy of atomic hydrogen on coronene cations (2 - 3.5 eV \citep{cazaux2016}) is deposited into the molecular system with every attachment event. At the same time, the system is subject to cooling by photon emission and fragmentation processes. The resulting internal excitation depends on the experimental conditions but only marginally on the initial temperature, as discussed elsewhere \citep{rapacioli2018}.

Molecular hydrogen or deuterium gas was dissociated in a Slevin-type discharge source operated at an RF frequency of 27 MHz \citep{hoekstra1991,boschman2012,reitsma2014}. The gas was cooled through collisions with a water-cooled sleeve and guided through Teflon tubes into the RF trap holding the C$_{24}$H$_{12}^+$ target. For both hydrogen and deuterium exposures, exposure times were set to 0.15, 0.5, 1, 3, 6, 9, 12, 15, and 40\;s in order to get broad view of the resulting superhydrogenation states. All hydrogenation and deuteration measurements were performed under otherwise identical experimental conditions.

Before the hydrogenation experiments, a reference measurement was made of the pristine coronene cation sample. The resulting mass spectrum can be found in the top panel of Figure 2. The main feature of the mass spectrum is the C$_{24}$H$_{12}^+$ precursor ion with $M=300$\;amu. A weak peak at $M=301$ is due to $^{12}$C$_{23}$$^{13}$C H$_{12}^+$.  This peak is due to the naturally occurring $^{13}$C isotope. In order to correct for this peak in the analysis of subsequent measurements, the ratio between the main coronene mass peak and this isotope peak was calculated. This ratio was used during calculations on all mass peaks in the subsequent measurements to correct for the presence of this isotope peak. It is of note that the isotopic contribution is much less than its natural fraction of almost 25\%, due to the mass filtering by the RF quadrupole. Tighter filtering starts to reduce the number of C$_{24}$H$_{12}^+$ cations of $M=300$ as well, thereby hampering performing experiments with sufficient statistics.

\begin{figure*}
\includegraphics[width=14cm]{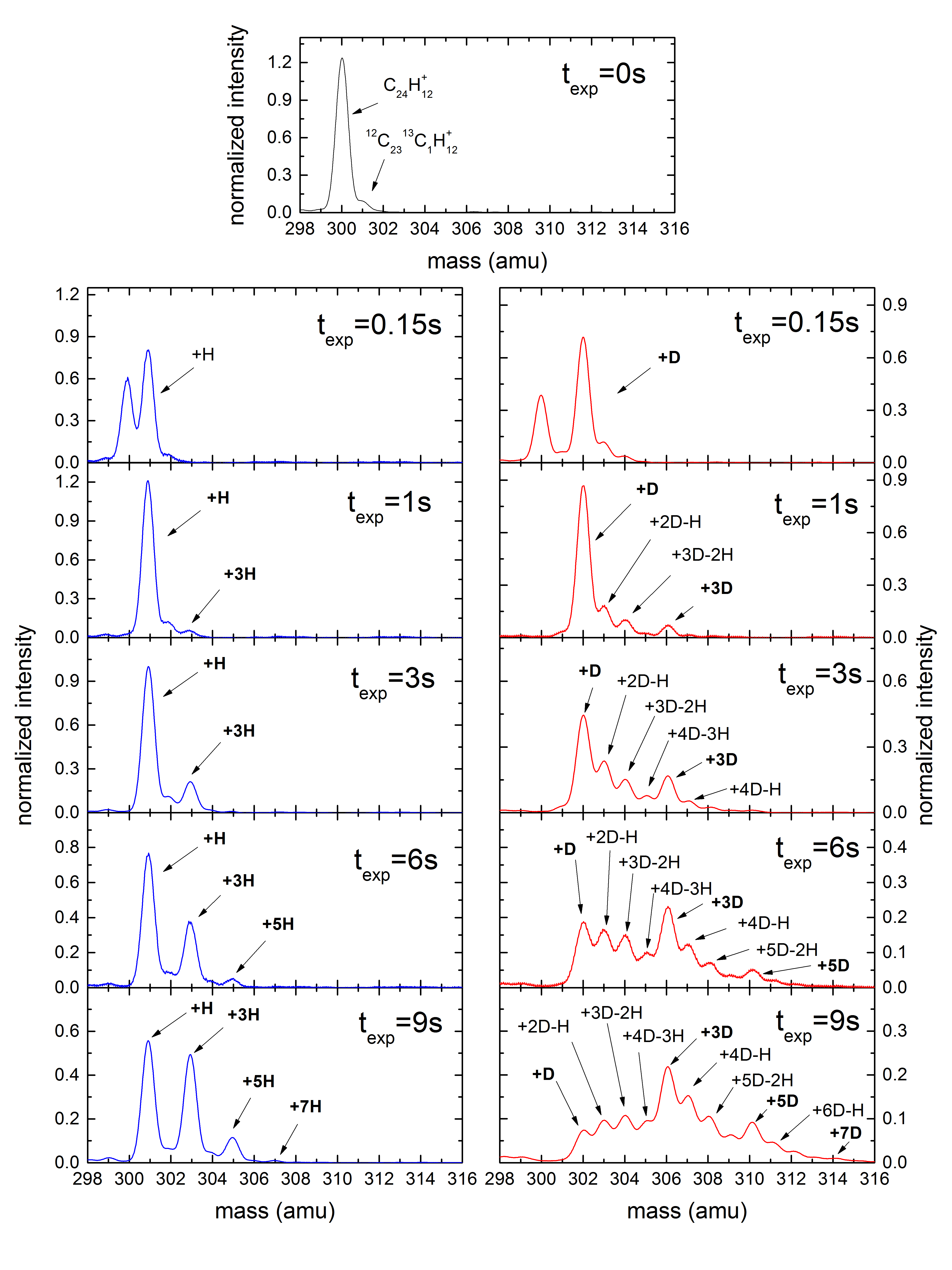}
\caption{Top panel: Mass spectrum of the coronene radical precursor C$_{24}$H$_{24}^+$ featuring a small contamination of $^{ 12}$C$_{23}$$^{13}$C$_1$H$_{24}^+$ (see text); Left column: Evolution of the superhydrogenation pattern as a function of H exposure time texp. Right column:  Evolution of the superhydrogenation pattern as a function of D exposure time t$_{exp}$. All distributions are normalized with respect to the total peak integral.}
\label{fig:spectra}
\end{figure*}

\subsection{Results}

The left panel in fig. \ref{fig:spectra} shows the evolution of the mass spectrum with increasing H exposure time. The pressure in the RF-trap chamber was at $p=1.5\times10^{-6}$ mbar.  Already after an H exposure time $t_{exp}=0.15$ s, more than half of the trapped ions are in the singly superhydrogenated state ($M=301$, +H). For $t_{exp}=1$ s, almost the entire trap content is singly superhydrogenated and a small feature due to triple superhydrogenation emerges ($M=303$, +3H). With increasing $t_{exp}$, the ratio between +H and +3H shifts towards the latter one and at $t_{exp}=6$ s, the next odd superhydrogenation appears ($M=305$, +5H). At an exposure time of $t_{exp}=9$ s, first traces of 7-fold superhydrogenation show up ($M=307$, +7H). In between the odd-mass peaks, the weak intensity at even mass numbers is mostly due to the presence of $^{13}$C in the precursor ion. The evolution towards the expected superhydrogenation pattern with its pronounced odd-even oscillation is evident. As discussed in the introduction, for atomic H exposure H abstraction remains hidden in the mass spectra as it leads to formation of molecular cations of identical masses.

The right panel of fig. \ref{fig:spectra} shows the corresponding results for D exposure at otherwise identical conditions. For the shortest exposure time of 0.15\ s, results for H and D exposure only differ in mass shift. In both cases, more than half of the C$_{24}$H$_{12}^+$ population has been transferred to the singly superhydrogenated or superdeuterated state. As discussed in the introduction, abstraction does not play a role in this step. 
For longer exposure times the deuteration spectra are much more complex than the hydrogenation ones. The generic trends, and especially the additional information contained in the more complex deuterium spectra, will be illustrated for the cases of 1 and 3\ s exposure.

For exposures times longer than 1 s, HD abstraction occurs, resulting in the appearance of peaks in between the ones corresponding to the attachment of 1 and 3 D atoms at $M=302$ and $M=306$, respectively. For 1 s exposure, peaks at $M=303$ and $M=304$ are clearly visible. As illustrated in figure \ref{fig:sketch}, the production of $M=303$ requires the subsequent attachment of 2 D atoms followed by a HD abstraction event. The peak at $M=304$ requires the attachment of 3 D atoms and 2 HD abstraction events.

For $t_{exp}=3$ s, hydrogenation of the coronene cations show singly hydrogenated species at $M=301$, as well as triply hydrogenated species at $M=303$. Deuteration of coronene cations, on the other hand, leads to the presence of singly and triply deuterated species at $M=302$ and $M=306$ respectively. However, many intermediate masses can be observed, which are resulting from abstraction reactions such as the masses $M=303, 304$, and 305 (+2D-H, +3D-2H, and +4D-3H, respectively). With further increase of $t_{exp}$, the hydrogenation pattern continues to evolve according to the same principle: new odd superdeuteration states appear (masses $M=310$, +5D) and subsequently become accompanied by higher mass species with higher D content and lower H content. Note that the labeling in the right panel of fig.\ref{fig:spectra} is not complete and only features the most straightforward sequences leading to odd total numbers of hydrogen and deuterium atoms involved. For longer exposure time of $t_{exp}=6$ s and $t_{exp}=9$ s, coronene cations with 5  and then with 7 extra deuterium atoms can be seen, as well as intermediate masses due to abstraction.

\begin{figure*}
\includegraphics[width=12cm]{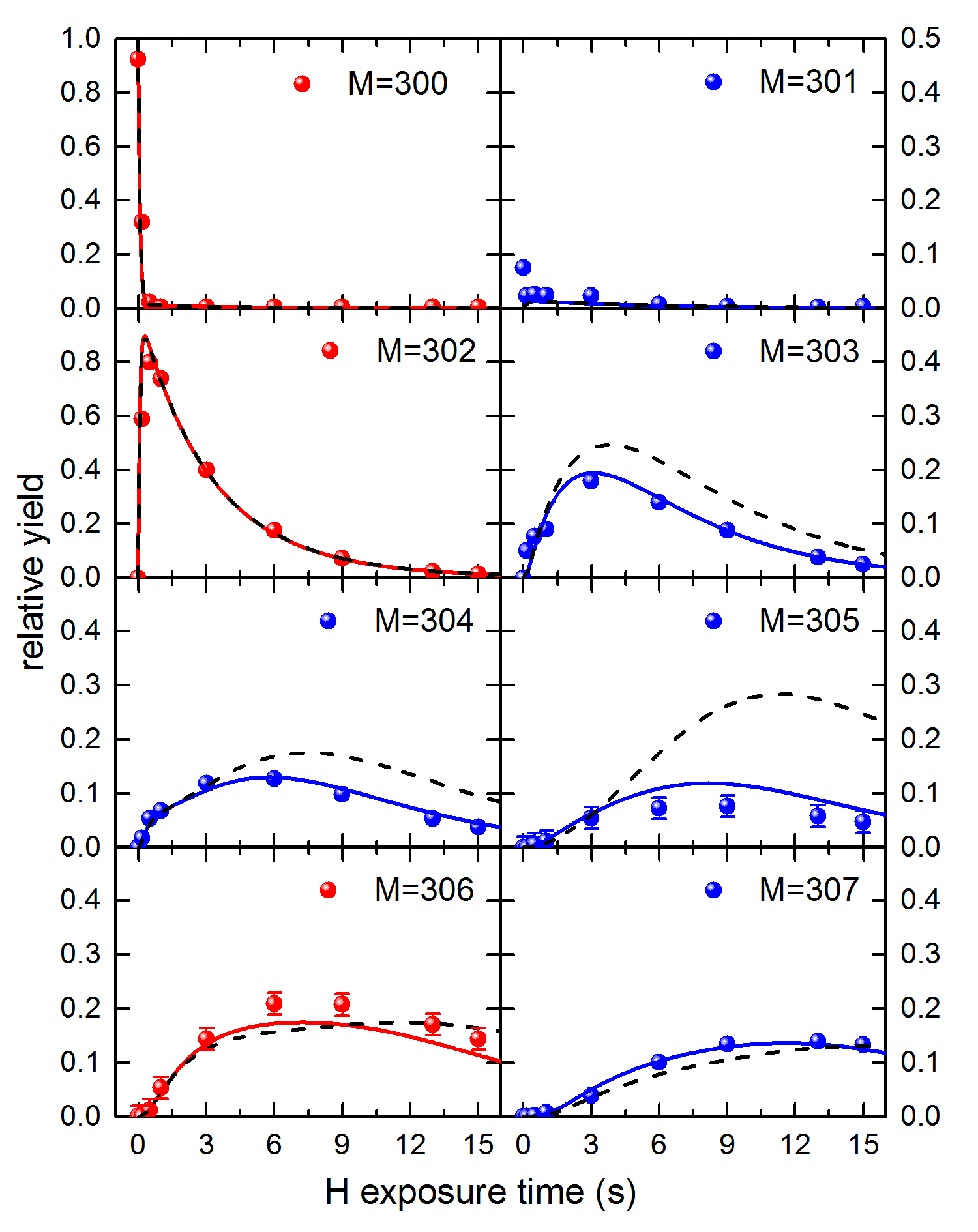}
\caption{Ion yields for mass numbers M=300-307 Da as a function of D exposure time. Solid data points: experimental data (red: masses resulting from pure D attachment; blue: masses that necessarily involve H abstraction). Dashed lines: Simulation results obtained with abstraction cross sections for graphene. Solid: modified abstraction barriers (see text). The experimental error bars reflect uncertainties 
from the data treatment: background subtraction, Gaussian fitting as well as the presence of a small contamination of 13C that we could not discriminate. }
\label{fig:deuterium_yields}
\end{figure*}

To conclude, for deuterated coronene the masses of stable superhydrogenation states follow the equation:
\begin{equation}
M_{\mbox{stable}}=300+2m-n, m-n=[1,3,5,]
\end{equation}
Here $m$ represents the number of D atoms attached to the molecule, and $n$ represents the number of H atoms that have been abstracted from the molecule.
For the next section it is important to realize that the composition of a stable superhydrogenation state defines the possible subsequent abstraction reactions. For instance, after HD abstraction from a deuterated site, a subsequent D attachment leaves the site doubly occupied with D atoms and accordingly, as a next step only D$_2$ abstraction is possible. In general the relative importance of D$_2$ abstraction will increase in the course of the attachment/abstraction sequence.

\section{Kinetic model}
\subsection{Assumptions}
In order to extract quantitative information on reaction cross sections and barriers from the experimental data, attachment and abstraction sequences were described with a rate-equation model. 
To this end, the time-evolution of each of the D-exposure peaks in the experimental data from fig. \ref{fig:spectra} up to $M=309$ was determined by integration of Gaussian fits as a function of $t_{exp}$. The data is shown in fig.~\ref{fig:deuterium_yields} as solid circles for each mass. To be able to reach $M=309$ (+4D -1H) not only by D-attachment, but also by HD/D$_2$-abstraction, attachment up to $M=312$ and abstraction needed to be taken into account. The precise reaction barriers and cross sections for attachment and abstraction are very likely site-dependent and could not be precisely derived from our model as this would involve too many free parameters. We chose to follow the sequence and scenario derived by experiments and DFT calculations from \citep{cazaux2016}. In that study, the barrier for addition of the first hydrogen had been estimated as $E_{attach}^{radical}=10$ meV, while the second hydrogenation had a higher barrier of $E_{attach}^{closed}=30$ meV, as it described a reaction between a closed shell cation and a radical. For the subsequent D attachment steps, we used this alternation between barriers of 10 meV and 30 meV until the 6$^{th}$ addition. For the 6$^{th}$ hydrogenation, a higher barrier of 100 meV had been determined. For abstraction processes, we used the small $E_{abstract}=10$ meV barriers determined for such reactions on graphene \citep{morisset2004}. Based on these assumptions, a chemical network capturing the different processes was established which is reported in the appendix.   

\subsection{Results}
We then compute the evolution of the yields of the different superdeuterated species using our chemical network. Our first goal is to reproduce experimental data using the values that were previously derived in other studies. First, we used cross sections for attachment (1.1 \AA$^2$) and abstraction (0.06 \AA$^2$) similar to the ones derived for neutral coronene by \citep{mennella2012} and independent on the hydrogenation state \cite{skov2014}. In this low abstraction scenario the computations do not reproduce the even mass-numbered superdeuterated states reported in fig.~\ref{fig:deuterium_yields}, which imply the abstraction of a H atom by a D to form HD. In order to reproduce the experimentally observed yields of all species whose formation involved abstraction processes, the abstraction cross sections had to be significantly increased to at least 0.45 \AA$^2$ while the addition cross sections are set to 0.9 \AA$^2$. This implies that for coronene cations, abstraction rates are at least half of addition rates. Note that since the flux of D atoms is difficult to constrain in our experiments, the cross sections derived in our model could be different. However, our model allows to derive the ratio between abstraction and addition cross sections.

Fig.~\ref{fig:deuterium_yields} displays the model results for the yields of various superdeuterated coronene cations as a function of their mass number $M$ as dashed lines.The agreement with the experimental data is good with the exception of the $M=303$ and $305$ case.

Our model gives a reliable ratio of the cross sections and thus of the reaction rates. It is important to realize that the absolute cross sections are related to the respective reaction barriers for attachment and abstraction. Barriers and cross sections cannot be determined independently with our approach. 
A closer look at the dashed lines in fig.~\ref{fig:deuterium_yields} reveals an overestimation of the yields for $M=303, 304$ and 305 whereas the yields for higher masses ($M=306, 307$) are underestimated. The two different attachment barriers $E_{attach}^{radical}=10$ meV ($n+m$ even, radical-radical reaction) and $E_{attach}^{closed}=30$ meV ($n+m$ odd, closed-shell-radical reaction) we used were theoretically determined for sequential hydrogenation in the absence of abstraction reactions \citep{cazaux2016}.
Abstraction reactions could induce a deviation from the hydrogenation sequence. 

For instance, sequential attachment of 3 D atoms to sites 1, 2 and 3 leads to formation of C$_{24}$H$_{12}$D$_3^+$ (see fig.\ref{fig:sketch}, D exposure to 12 H, $n=0$ , $M=306$). Starting from this configuration, an abstraction could remove an H from one of the neighboring doubly occupied outer-edge sites, e.g. site 2 (see fig.\ref{fig:sketch}, the molecule shifts down-left to 11 H, $n=1$, $M=305$). The result is a radical configuration that would not be formed by attachment processes only, with a single doubly occupied outer edge site 1 and a singly occupied inner edge site 3. Attachment barriers typically decrease upon structural perturbations \citep{cazaux2016} which could result in attachment of the next D atom not only to site 2 but also to site 4. The resulting closed shell cation would have either a 1, 2, 3 conformation (with the correspondingly high attachment barrier $E_{attach}^{closed}=30$) or it would have a 1, 3, 4 configuration. In the latter case, the attachment barrier could be lowered, because there are now two singly occupied outer edge sites with a perturbed molecular structure due to neighboring doubly occupied outer edge sites. Similar reasoning is not limited to an abstraction from the triply superhydrogenated configuration but to all configurations that have originally undergone  $n\geq 2$ D attachment processes and at least one abstraction reaction. 

Abstraction reactions could also reduce the barrier for subsequent attachment, because abstraction leads to extra vibrational excitation of the remaining coronene cation. For similar Eley-Rideal processes on graphite, it has been theoretically shown that most of the released energy goes into the vibrational energy of the substrate rather than into the formed H$_2$ \citep{bachellerie2007,sizun2010}. On the other hand, attachment also increases vibrational excitation and does not lead to decreasing attachment barriers.

We have tried to reflect this with our model, by lowering the  attachment barriers $E_{attach}^{closed}$ from 30 meV to 10 meV for the discussed closed shell configurations ($n\geq 2$, at least one abstraction, see appendix barriers in bold face).
The resulting yields are shown in fig.\ref{fig:deuterium_yields} as solid lines. The increased agreement between model results and experimental data is obvious. 

While our model reproduces experimental data better when attachment barriers for all configurations modified by abstraction reactions are reduced, we would like to stress that other mechanisms cannot be ruled out. The number of potential attachment sites could be different for different configurations corresponding to the same hydrogenation state. Also, abstraction reactions could lead to an increase in internal energy.

\section{Conclusions}
The main conclusion from our study is therefore the 2:1 ratio between attachment and abstraction.  This implies that H abstraction from gas phase coronene cations is at least more than 7 times more efficient than H abstraction from  neutral coronene thin films. This could have implications for H$_2$ production in the ISM. For instance, Andrews {\it et al.} \citep{andrews2016} have recently shown that in photodissociation regions, PAHs only contribute to H$_2$ formation via photodissociation channels and not via abstraction mechanisms. However, their calculations were based on the low abstraction cross sections from \citep{mennella2012}. An increase of the abstraction cross section by one order of magnitude could make PAHs a very important route for the formation of H$_2$ in space.

\section*{Acknowledgements}
\bibliographystyle{mnras}

\appendix
\section*{Supplementary information}
We used a simple rate equation approach to compute the deuterium addition and abstraction on/from coronene cations. The different reactions considered in our chemical network are shown in table \ref{tab:barriers2}. Each reaction rate is computed as R=$g\times \sigma \times \exp({\frac{-Ea}{T}})$, where g is the degeneracy, Ea the reaction barrier and sigma the cross section. The formation of a species x through the reaction of y and z is described by $\frac{dn_x}{dt}= R * n_y n_z$, while the destruction of the species x through reaction with z is $\frac{dn_x}{dt}= - R * n_x n_z$. The addition and abstraction cross sections reproducing the experimental measurements are listed in the table. 
\begin{table*}
	\centering
    \small
	\caption{Cross sections, barriers and degeneracies for the various attachment and abstraction channels as used for the numerical modeling leading to the solid lines in fig.\ref{fig:deuterium_yields}.}
    \label{tab:barriers2}
	\scalebox{0.8}{
    \begin{tabular}{|cc|cc|ccccc} 
    \hline
    initial & incoming & final & abstraction & cross& reaction & degeneracy & initial & final \\
    PAH & atom & PAH & product & section (\AA$^2$)& barrier & & mass & mass\\
		\hline
Cor$^+$     &D &    CorD$^+$     &      &0.9 &0.010    &12   &300&302\\
CorD$^+$    &D &     Cor$^+$     &D2  &0.45&0.010    &1&302&300\\
CorD$^+$     &D&   CorD$_2^+$      &   &0.9 &0.030    &1&302&304\\
CorD$^+$     &D&   (CorD-H)$^+$ &HD  &0.45&0.010    &1&302&301\\
(CorD-H)$^+$ &D&   (CorD$_2$-H)$^+$  &    &0.9 &0.010    &12   &301&303\\
CorD$_2^+$      &D&   CorD$^+$    &D$_2$  &0.45&0.010  &  2&304&302\\
CorD$_2^+$      &D&   CorD$_3^+$    &    &0.9 &0.010  &  4&304&306\\
CorD$_2^+$      &D&   (CorD$_2$-H)$^+$   &HD  &0.45&0.010    &2&304&303\\
(CorD$_2$-H)$^+$    &D&   (CorD-H)$^+$  &D$_2$  &0.45&0.010    &1&303&301\\
(CorD$_2$-H)$^+$   &D&   (CorD$_3$-H)$^+$  &      &0.9 &\bf{0.030/0.010}    &1&303&305\\
(CorD$_2$-H)$^+$    &D&   (CorD$_2$-H$_2$)$^+$  &HD  &0.45&0.010    &1&303&302\\
(CorD$_2$-H$_2$)$^+$  &D&   (CorD$_3$-H$_2$)$^+$ &      &0.9 &0.010    &12   &302&304\\
CorD$_3^+$   &D&   CorD$_2^+$   &D$_2$  &0.45&0.010    &1&306&304\\
CorD$_3^+$   &D&   CorD$_4^+$    &      &0.9 &0.030    &1&306&308\\
CorD$_3^+$   &D&   (CorD$_3$-H)$^+$   &HD  &0.45&0.010    &1&306&305\\
(CorD$_3$-H)$^+$ &D&   (CorD2-H)$^+$ &D$_2$  &0.45&0.010    &2&305&303\\
(CorD$_3$-H)$^+$ &D&   (CorD$_4$-H)$^+$ &       &0.9 &0.010    &4&305&307\\
(CorD$_3$-H)$^+$ &D&   (CorD$_3$-H$_2$)$^+$ &HD  &0.45&0.010    &2&305&304\\
(CorD$_3$-H$_2$)$^+$&D&   (CorD2-H$_2$)$^+$ &D$_2$  &0.45&0.010    &1&304&302\\
(CorD$_3$-H$_2$)$^+$&D&   (CorD$_4$-H$_2$)$^+$ &      &0.9 &\bf{0.030/0.010}    &1&304&306\\
(CorD$_3$-H$_2$)$^+$&D&  (CorD$_3$-H$_3$)$^+$ &HD  &0.45&0.010    &1&304&303\\
(CorD$_3$-H$_3$)$^+$&D&   (CorD$_4$-H$_3$)$^+$  &     &0.9 &0.010    &12   &303&305\\
CorD$_4^+$   &D&   CorD3$^+$   &D$_2$  &0.45&0.010    &4&308&306\\
CorD$_4^+$   &D&   CorD5$^+$    &      &0.9 &0.010    &1    &308    &310\\
CorD$_4^+$   &D   &   (CorD$_4$-H)$^+$   &HD      &0.45    &0.010    &4    &308    &307\\
(CorD$_4$-H)$^+$ &D   &   (CorD$_3$-H)$^+$ &D$_2$      &0.45    &0.010    &3    &307    &305\\
(CorD$_4$-H)$^+$ &D   &   (CorD$_5$-H)$^+$ &          &0.9     &\bf{0.030/0.010}    &1    &307    &309\\
(CorD$_4$-H)$^+$ &D   &   (CorD$_4$-H$_2$)$^+$  &HD      &0.45    &0.010    &3    &307    &306\\
(CorD$_4$-H$_2$)$^+$&D   &   (CorD$_3$-H$_2$)$^+$&D$_2$      &0.45    &0.010    &2    &306    &304\\
(CorD$_4$-H$_2$)$^+$&D   &   (CorD5-H$_2$)$^+$  &         &0.9     &0.010    &4    &306    &308\\
(CorD$_4$-H$_2$)$^+$&D   &   (CorD$_4$-H$_3$)$^+$ &HD      &0.45    &0.010    &2    &305    &303\\
(CorD$_4$-H$_3$)$^+$&D   &   (CorD$_3$-H$_3$)$^+$ &D$_2$      &0.45    &0.010    &1    &305    &303\\
(CorD$_4$-H$_3$)$^+$&D   &   (CorD$_5$-H$_3$)$^+$  &         &0.9     &\bf{0.030/0.010}    &1    &305    &307\\
(CorD$_4$-H$_3$)$^+$&D   &   (CorD$_4$-H$_4$)$^+$ &HD      &0.45    &0.010    &1    &305    &304\\
(CorD$_4$-H$_4$)$^+$&D   &   (CorD$_5$-H$_4$)$^+$  &         &0.9     &0.010    &12   &304    &306\\
CorD$_5^+$   &D   &   CorD$_4^+$   &D$_2$      &0.45    &0.010    &   5    &310    &308\\
CorD$_5^+$   &D   &   CorD$_6^+$    &          &0.9     &0.100    &4    &310    &312\\
CorD$_5^+$   &D   &   (CorD$_5$-H)$^+$  &HD      &0.45    &0.010   &    5    &310    &309\\
(CorD$_5$-H)$^+$ &D   &   (CorD$_4$-H)$^+$ &D$_2$      &0.45    &0.010    &4    &309    &307\\
(CorD$_5$-H)$^+$ &D   &   (CorD$_6$-H)$^+$  &          &0.9     &0.010    &1    &309    &311\\
(CorD$_5$-H)$^+$ &D   &   (CorD5-H$_2$)$^+$ &HD      &0.45    &0.010    &4    &309    &308\\
(CorD$_5$-H$_2$)$^+$&D   &  (CorD$_4$-H$_2$)$^+$&D$_2$      &0.45    &0.010    &3    &308    &306\\
(CorD$_5$-H$_2$)$^+$&D   &   (CorD$_6$-H2)$^+$  &         &0.9     &0.030    &1    &308    &310\\
(CorD$_5$-H$_2$)$^+$&D   &   (CorD$_5$-H$_3$)$^+$ &HD      &0.45    &0.010    &3    &308    &307\\
(CorD$_5$-H$_3$)$^+$&D   &   (CorD$_4$-H$_3$)$^+$ &D$_2$      &0.45    &0.010    &2    &307    &305\\
(CorD$_5$-H$_3$)$^+$&D   &   (CorD$_6$-H$_3$)$^+$ &          &0.9     &0.010    &4    &307    &309\\
(CorD$_5$-H$_3$)$^+$&D   &   (CorD$_5$-H$_4$)$^+$ &HD      &0.45    &0.010    &2    &307    &306\\
(CorD$_5$-H$_4$)$^+$&D   &   (CorD$_4$-H$_4$)$^+$ &D$_2$      &0.45    &0.010    &1    &306    &304\\
(CorD$_5$-H$_4$)$^+$&D   &   (CorD$_6$-H$_4$)$^+$ &          &0.9     &0.030    &1    &306    &308\\
(CorD$_5$-H$_4$)$^+$&D   &   (CorD$_5$-H$_5$)$^+$ &HD      &0.45    &0.010    &1    &306    &305\\
(CorD$_5$-H$_5$)$^+$&D   &   (CorD$_6$-H$_5$p &          &0.9     &0.010    &12   &305    &307\\
CorD$_6^+$   &D   &   CorD$_5^+$   &D$_2$      &0.45    &0.010  &     6    &312    &310\\
CorD$_6^+$   &D   &   CorD$_7^+$   &         &0.9     &0.010    &1    &312    &314\\
CorD$_6^+$   &D   &   (CorD$_6$-H)$^+$   &HD      &0.45    &0.010    & 6    &312    &311\\
(CorD$_6$-H)$^+$ &D   &   (CorD$_5$-H)$^+$ &D$_2$      &0.45    &0.010    &   5    &311    &309\\
(CorD$_6$-H)$^+$ &D   &   (CorD$_7$-H)$^+$ &         &0.9     &0.100    &1    &311    &313\\
(CorD$_6$-H)$^+$ &D   &   (CorD$_6$-H$_2$)$^+$  &HD      &0.45    &0.010    &   5    &311    &310\\
(CorD$_6$-H$_2$)$^+$&D   &   (CorD$_5$-H$_2$)$^+$&D$_2$      &0.45    &0.010    &4    &310    &308\\
(CorD$_6$-H$_2$)$^+$&D   &   (CorD$_7$-H$_2$)$^+$  &         &0.9     &0.010    &1    &310    &312\\
(CorD$_6$-H$_2$)$^+$&D   &   (CorD$_6$-H$_3$)$^+$ &HD      &0.45    &0.010    &4    &310    &309\\
(CorD$_6$-H$_3$)$^+$&D   &   (CorD$_5$-H$_3$)$^+$ &D$_2$      &0.45    &0.010    &3    &309    &307\\
(CorD$_6$-H$_3$)$^+$&D   &   (CorD$_7$-H$_3$)$^+$ &          &0.9     &0.030    &1    &309    &311\\
(CorD$_6$-H$_3$)$^+$&D   &   (CorD$_6$-H4)$^+$ &HD      &0.45    &0.010    &3    &309    &308\\
(CorD$_6$-H$_4$)$^+$&D   &   (CorD$_5$-H4)$^+$ &D$_2$      &0.45    &0.010    &4    &308    &306\\
(CorD$_6$-H$_4$)$^+$&D   &   (CorD$_7$-H4)$^+$ &          &0.9     &0.010    &2    &308    &310\\
(CorD$_6$-H$_4$)$^+$&D   &   (CorD$_6$-H$_5$)$^+$ &HD      &0.45    &0.010    &4    &308    &307\\
(CorD$_6$-H$_5$)$^+$&D   &   (CorD5-H$_5$)$^+$ &D$_2$      &0.45    &0.010    &1    &307    &305\\
(CorD$_6$-H$_5$)$^+$&D   &   (CorD$_7$-H$_5$)$^+$ &          &0.9     &0.030    &1    &307    &309\\
(CorD$_6$-H$_5$)$^+$&D   &   (CorD$_6$-H$_6$)$^+$ &HD      &0.45    &0.010    &1    &307    &306\\
(CorD$_6$-H$_6$)$^+$&D   &   (CorD$_7$-H$_6$)$^+$  &         &0.9     &0.010    &12   &306    &308\\
		\hline
	\end{tabular}}
\end{table*}

\label{lastpage}

\end{document}